\documentclass[final,1p,times,twocolumn]{elsarticle}
\usepackage{lineno,hyperref,amsmath,amsthm,amssymb,amsfonts,ragged2e,color,subfig}
\usepackage[flushleft]{threeparttable}
\usepackage{booktabs,caption}
\journal{``Chinese Journal of Physics"}
\begin{document}
\begin{frontmatter}
\title{Ion-acoustic rogue waves in a multi-component plasma medium}
\author{S. Jannat$^{*1}$,  N. A. Chowdhury$^{**,1,2}$,  A. Mannan$^{\dag,1,3}$, and A. A. Mamun$^{\ddag,1}$}
\address{$^1$Department of Physics, Jahangirnagar University, Savar, Dhaka-1342, Bangladesh\\
$^2$Plasma Physics Division, Atomic Energy Centre, Dhaka-1000, Bangladesh\\
$^3$Institut f{\"u}r Mathematik, Martin Luther Universit{\"a}t Halle-Wittenberg, D-06099 Halle, Germany\\
e-mail: $^*$jannat23phy@gmail.com, $^{**}$nurealam1743phy@gmail.com,\\
$^{\dag}$abdulmannan@juniv.edu, $^\ddag$mamun\_phys@juniv.edu}
\begin{abstract}
The nonlinear propagation of ion-acoustic (IA) waves (IAWs) in a four component plasma
medium (FCPM) containing inertial warm positive ions, and inertialess iso-thermal
cold electrons as well as non-extensive ($q$-distributed) hot electrons and positrons
is theoretically investigated. A nonlinear Schr\"{o}dinger equation (NLSE) is derived
by using the reductive perturbation method, and it is observed that the FCPM under consideration supports
both modulationally stable and unstable parametric regimes which are determined by the
sign of the dispersive and nonlinear coefficients of NLSE. The numerical analysis has
shown that the maximum value of the growth rate decreases with the increase in $q$ ($q>1$),
and the modulationally unstable parametric regime
allows to generate highly energetic IA rogue waves (IARWs), and the amplitude and width of the IARWs
increase with an increase in the value of hot electron number density while decrease with
an increase in the value of cold electron number density.
The applications of our investigation in understanding the basic features of nonlinear electrostatic
perturbations in many space plasma environments and laboratory devices are briefly discussed.
\end{abstract}
\begin{keyword}
NLSE \sep Modulational instability \sep Ion-acoustic waves \sep Rogue waves.
\end{keyword}
\end{frontmatter}
\section{Introduction}
\label{3sec:Introduction} The existence of the
electron-positron-ion (EPI) plasma has been identified in astrophysical environments
such as Saturn's magnetosphere \cite{Rehman2016,Shahmansouri2013,Shalini2015,Panwar2014,Kourakis2003,Alinejad2014},
pulsar magnetosphere, active galactic nuclei, early universe \cite{Rehman2016}, neutron stars,
Sun atmosphere \cite{Rehman2016}, and has also been confirmed in various laboratory experiments
such as intense laser field. The formation and propagation of various kinds of electrostatic
waves namely, ion-acoustic (IA) waves  (IAWs) \cite{Rehman2016,Shahmansouri2013,Shalini2015,Panwar2014,Kourakis2003,Alinejad2014,Baluku2012},
electron-acoustic waves (EAWs) \cite{Baluku2011}, and positron-acoustic waves (PAWs) as well as their
associated nonlinear structures (viz., solitons, double layers, shocks, and vortices, etc.)
in space and laboratory EPI plasma have been significantly modified by the presence of positrons.

The co-existence of hot and cold electrons in Saturn's magnetosphere has
been identified by the Voyager PLS observations \cite{Schippers2008,Sittler1983,Barbosa1993} and
the CAPS (Cassini Plasma Spectrometer) observations \cite{Young2005}, and
this identification has attracted a number of authors
\cite{Rehman2016,Shahmansouri2013,Shalini2015,Panwar2014,Kourakis2003,Alinejad2014,Baluku2012}
to study the nonlinear properties of the plasma system having two
temperature electrons. Rehman and Mishra \cite{Rehman2016}
analytically and numerically analyzed the IA Gardner solitons in
an EPI plasma with two temperature electrons. Shahmansouri and Alinejad
\cite{Shahmansouri2013} studied IA solitary waves in an three
components plasma medium having inertialess cold and hot electrons
as well as inertial ions, and reported that this model supports
both compressive and rarefactive solitary structures in presence
of two temperature electrons.   Panwar \textit{et al.} \cite{Panwar2014}
considered three components plasma model having inertial positive
ion and inertialess cold and hot electrons to study the
propagation of nonlinear IA cnoidal waves, and found that the
height and width of a cnoidal waves increase with the ratio of
cold and hot electrons temperature. Baluku and Helberg
\cite{Baluku2012} examined the nonlinear properties of the IA
solitons in a plasma with two temperature electrons.

The deviation of the plasma species from the equilibrium state
due to the activation of long range coulomb force field, wave-particle
interaction, and other external force fields is
described by the non-extensive $q$-distribution \cite{Shalini2015}. The parameter
$q$ in $q$-distribution indicates the non-extensive
properties of the plasma species in a non-equilibrium plasma
system, and when $q$ is equal to unity then the $q$-distribution
coincides with well-known Maxwellian distribution. When
$q$ is less than one (i.e., $q<1$) then the plasma species
indicate the super-extensive properties while $q$ is greater than
one (i.e., $q>1$) then the plasma species indicate the
sub-extensive properties \cite{Shalini2015}.

\begin{figure}[t!]
\centering
\includegraphics[width=70mm]{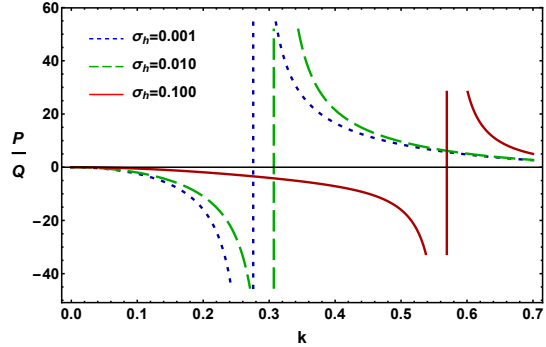}
\caption{The variation of $P/Q$ with $k$ for different values of
$\sigma_h$ when $\alpha$=$0.04$, $\gamma_c=0.5$, $\gamma_h=0.8$,  $\sigma_p=0.01$, $q_h=1.3$, and $q_p=1.5$.}
\label{3Fig:F1}
\end{figure}
The standard nonlinear Schr\"{o}dinger
equation (NLSE) is the first platform to investigate the
nonlinear properties of the dispersive plasma medium as well as
the modulational instability (MI) of IAWs, EAWs, and PAWs as well as their
associated first and second order rogue waves (RWs) in
the dispersive plasma medium \cite{Ahmed2018,Khondaker2019,Chowdhury2017,Rahman2018a,Jahan2019,Rahman2018b,Chowdhury2019}.
 Shalini \textit{et al.} \cite{Shalini2015} considered
three component plasma model having inertialess two temperature
electrons and inertial ions, and studied the MI of IAWs, and
observed the effects of the non-extensivity of hot and cold
electrons. Kourakis and Shukla \cite{Kourakis2003} considered three
component plasma model in presence of two temperature electron
species, and demonstrated the MI of IAWs by deriving standard
NLSE, and reported that a strong temperature difference between
hot and cold electrons may favourable to bright envelope solitons.
Alinejad \textit{et al.} \cite{Alinejad2014} studied the
stability of the IAWs in a plasma medium having two temperature
electrons. Therefore, in our present paper, we will study the MI
of the IAWs and the mechanism of generating the first and second order
IA RWs (IARWs) in a four component plasma medium (FCPM) having inertial warm ions,
and inertialess iso-thermal cold electrons and non-extensive hot electrons and positrons.

The manuscript is organized as follows: The basic model equations
are presented in Sec. \ref{3sec:Governing Equations}. A NLSE is derived in
Sec. \ref{3sec:Derivation of NLSE}. The MI and RWs are
provided in Sec. \ref{3sec:Modulational instability and Rogue waves}.
Results and discussion are presented in Sec. \ref{3sec:Results and discussion}.
Finally, a brief conclusion is given in Sec. \ref{3sec:Conclusion}.
\section{Governing Equations}
\label{3sec:Governing Equations}
We consider a four component unmagnetized plasma model consisting
of inertial warm ions, inertialess non-extensive hot electrons and
positrons as well as iso-thermal cold electrons following Maxwellian
distribution. At equilibrium, the overall charge neutrality condition
for our plasma model can be written as $ Z_i n_{i0}+n_{p0} = n_{c0}+
n_{h0}$; where $n_{i0}$, $n_{p0}$, $n_{c0}$, and $n_{h0}$ are the
equilibrium number densities of warm ions, non-extensive positrons,
and cold and hot electrons, respectively, and $Z_i$  is the number of
protons residing onto the ion surface. The normalized governing
equations to study the IAWs  are as follows:
\begin{eqnarray}
&&\hspace*{-1.3cm}\frac{\partial n_i}{\partial t}+\frac{\partial}{\partial x}(n_{i} u_{i})=0,
\label{3eq:1}\\
&&\hspace*{-1.3cm}\frac{\partial u_i}{\partial t}+u_{i}\frac{\partial u_i}{\partial x}+ \alpha n_{i}\frac{\partial n_i}{\partial x}= - \frac{\partial\phi}{\partial x},
\label{3eq:2}\\
&&\hspace*{-1.3cm}\frac{\partial^2 \phi}{\partial x^2}=\gamma_cn_c+\gamma_h n_h-(\gamma_c+\gamma_h-1) n_p-n_i,
\label{3eq:3}\
\end{eqnarray}
where  $n_i$  is the number density of inertial warm ions normalized by its equilibrium value $n_{i0}$;
$u_i$ is the ion fluid speed normalized by the IAW speed $C_i=(Z_ik_BT_c/m_i)^{1/2}$ (with $T_c$
being the $q$-distributed cold electron temperature, $m_i$ being the ion rest mass, and $k_B$
being the Boltzmann constant); $\phi$ is the electrostatic wave potential normalized by $k_BT_c/e$
(with $e$ being the magnitude of single electron charge); the time and space variables are
normalized by ${\omega^{-1}_{pi}}=(m_i/4\pi Z^2 e^2 n_{i0})^{1/2}$ and $\lambda_{Di}=(k_BT_c/4 \pi Z e^2
n_{i0})^{1/2}$, respectively. The pressure term of the ion  can be written as
$p_i=p_{i0}(N_{i}/n_{i0})^\gamma$ with $p_{i0}=n_{i0}k_BT_i$ being the equilibrium  pressure of the
ion, and $T_i$ being the temperature of warm ion,  and $\gamma=(N+2)/N$ (where $N$ is the degrees of freedom and
for one-dimensional case $N=1$ then $\gamma=3$). Other parameters are defined as $\alpha=3T_i/Z_iT_c$,
$\gamma_c=n_{c0}/Z_in_{i0}$, and $\gamma_h=n_{h0}/Z_in_{i0}$. The expression for the number density
of cold electrons following the Maxwellian distribution can be expressed as
\begin{eqnarray}
&&\hspace*{-1.3cm}n_c=\mbox{exp}(\phi)=1+\phi+\frac{\phi^2}{2}+\frac{\phi^3}{6}+\cdot\cdot\cdot,
\label{3eq:4}\
\end{eqnarray}
Now, the expressions for the number density of hot electrons
following the $q$-distribution can be expressed as \cite{Chowdhury2017}
\begin{eqnarray}
&&\hspace*{-1.3cm}n_h=[1+(q_h-1)\sigma_h\phi]^{\frac{(q_h+1)}{2(q_h-1)}}=1+A_1\phi+A_2\phi^2+A_3\phi^3+\cdot\cdot\cdot,
\label{3eq:5}\
\end{eqnarray}
where $q_h$ is the non-extensivity of the hot electrons, $\sigma_h$ = ${T_c}/{T_h}$ (with $T_h$ being the $q$-distributed hot electron temperature), and
\begin{eqnarray}
&&\hspace*{-1.0cm}A_1=[(q_h+1)\sigma_h]/2,~~~A_2=[(q_h+1)(3-q_h)\sigma_h^2]/8,~~~A_3=[(q_h+1)(3-q_h)(5-3q_h)\sigma_h^3]/48.
\nonumber\
\end{eqnarray}
Now, the expressions for the  number density of hot positrons following the $q$-distribution can be
expressed as \cite{Chowdhury2017}
\begin{eqnarray}
&&\hspace*{-1.3cm}n_p=[1-(q_p-1)\sigma_p\phi]^{\frac{(q_p+1)}{2(q_p-1)}}=1-A_4\phi+A_5\phi^2-A_6\phi^3+\cdot\cdot\cdot,
\label{3eq:6}\
\end{eqnarray}
where $q_p$ is the non-extensivity of the positrons, $\sigma_p$ = ${T_c}/{T_p}$ (with $T_p$ being the $q$-distributed positron temperature), and
\begin{eqnarray}
&&\hspace*{-0.7cm}A_4=[(q_p+1)\sigma_p]/2,~~~A_5=[(q_p+1)(3-q_p)\sigma_p^2]/8,~~~A_6=[(q_p+1)(3-q_p)(5-3q_p)\sigma_p^3]/48.
\nonumber\
\end{eqnarray}
The parameter $q_{h}$ and $q_{p}$ are generally known as entropic index.
Now, by substituting Eqs. \eqref{3eq:4}-\eqref{3eq:6} into Eq. \eqref{3eq:3} and expanding up to third order of $\phi$, we get
\begin{eqnarray}
&&\hspace*{-1.3cm}\frac{\partial^2\phi}{\partial x^2}+n_i=1+A_7\phi+A_8\phi^2+A_9\phi^3+\cdot\cdot\cdot,
\label{3eq:7}
\end{eqnarray}
where
\begin{eqnarray}
&&\hspace*{-1.3cm}A_7=\gamma_c+\gamma_hA_1+(\gamma_c+\gamma_h-1)A_4,
\nonumber\\
&&\hspace*{-1.3cm}A_8=[\gamma_c+{2\gamma_hA_2-2(\gamma_c+\gamma_h-1)A_5}]/2,
\nonumber\\
&&\hspace*{-1.3cm}A_9 =[\gamma_c+{6\gamma_hA_3+6(\gamma_c+\gamma_h-1)A_6}]/6.
\nonumber\
\end{eqnarray}
We note that Eq. \eqref{3eq:1},  \eqref{3eq:2}, and  \eqref{3eq:7} now represent the basis set of
normalized equations to describe the nonlinear dynamics of the IAWs, and associated IARWs in
the FCPM under consideration. We also note that the
works \cite{Ahmed2018,Khondaker2019,Chowdhury2017,Rahman2018a,Jahan2019,Rahman2018b,Chowdhury2019}
may seem to be similar to our present investigation,
but, in fact, they are not due to the following reasons:
\begin{itemize}
  \item Ahmed \textit{et al.} \cite{Ahmed2018}, Khondaker \textit{et al.} \cite{Khondaker2019}, and
  Chowdhury \textit{et al.} \cite{Chowdhury2017} studied
  the MI of IAWs, in which the moment of inertial is provided by the positive and negative ions and the
  restoring force is provided by the thermal pressure of the non-thermal (super-thermal $\kappa$-distributed
  and  $q$-distributed) electrons and positrons in a pair-ion plasma medium, and observed the existence of
  the IARWs in the modulationally unstable parametric regimes. However, in our present work we
  have considered a FCPM consisting of inertial warm ions, $q$-distributed positrons, and non-inertial
  two temperatures electrons [say, hot electrons (following $q$-distribution), cold electrons (following
  Maxwellian distribution)]. We have examined the conditions of MI of the IAWs (where, the warm positive
  ions provides the moment of inertia and the thermal pressure of the positrons and two temperature
  electrons provides the restoring force).
  \item Rahman \textit{et al.} \cite{Rahman2018a} and Jahan \textit{et al.} \cite{Jahan2019}
  investigated the stable and unstable parametric regimes of the dust-acoustic waves (DAWs) according
  to the sign of dispersive and nonlinear coefficients of the standard NLSE in a FCPM
  having inertial opposite polarity dust grains and inertialess non-thermal or iso-thermal
  ions as well as non-extensive electrons. Rahman \textit{et al.} \cite{Rahman2018b} analyzed theoretically and numerically
  the MI conditions of the DAWs in a FCPM having inertial cold and hot dust grains and  inertialess
  non-extensive electrons and ions. But our present work is concerned with the MI of IAWs in presence of hot and cold electron
  species.
  \item Chowdhury \textit{et al.} \cite{Chowdhury2019} studied the formation of only first-order IARWs in a FCPM
  having inertial positive ions and inertialess iso-thermal positrons as well as two temperature (hot and cold) electrons
  featuring super-thermal $\kappa$-distribution. On the other hand, in our present work, we have considered a FCPM
  consisting inertial positive ions and inertialess iso-thermal cold electrons as well as hot electrons and
  positrons featuring non-extensive $q$-distribution for studying the MI of IAWs and the mechanism of formation of the first
  and second order IARWs in the modulationally unstable parametric regime. It is important to mention here that the distribution
  function of the existing fast particles in any plasma medium is an important factor for developing the nonlinear properties of the
  plasma medium. So, the existence of $\kappa$-distributed or $q$-distributed  particles in a plasma medium rigorously changes the
  dynamics of that plasma medium, and the effects of $\kappa$-distributed particles are not similar with $q$-distributed particles.
\end{itemize}
\section{Derivation of the NLSE}
\label{3sec:Derivation of NLSE}
To study the MI of the IAWs, we will derive the NLSE by employing the reductive perturbation method. So, we
first introduce the stretched co-ordinates \cite{Chowdhury2019,Kourakis2005,Sultana2011,Schamel2002,Fedele2002}
\begin{eqnarray}
&&\hspace*{-1.3cm}\xi={\epsilon}(x-V_g t),
\label{3eq:8}\\
&&\hspace*{-1.3cm}\tau={\epsilon}^2 t, \label{3eq:9}
\end{eqnarray}
where $V_g$ is the group speed and $\epsilon$ is a small parameter measuring the strength of the
wave amplitude. Then we can write the dependent variables as
\cite{Chowdhury2019,Kourakis2005,Sultana2011,Schamel2002,Fedele2002}
\begin{eqnarray}
&&\hspace*{-1.3cm}n_i=1+\sum_{m=1}^{\infty}\epsilon^{m}\sum_{l=-\infty}^{\infty}n_{il}^{(m)}(\xi,\tau)~\mbox{exp}[il(kx-\omega
t)], \label{3eq:10}\\
&&\hspace*{-1.3cm}u_i=\sum_{m=1}^{\infty}\epsilon^{m}\sum_{l=-\infty}^{\infty}u_{il}^{(m)}(\xi,\tau)~\mbox{exp}[il(kx-\omega
t)], \label{3eq:11}\\
&&\hspace*{-1.3cm}\phi=\sum_{m=1}^{\infty}\epsilon^{m}\sum_{l=-\infty}^{\infty}\phi_l^{(m)}(\xi,\tau)~\mbox{exp}[il(kx-\omega
t)], \label{3eq:12}\
\end{eqnarray}
where $k$ ($\omega$) is real variable representing the carrier wave number (frequency).
The derivative operators in the above equations are treated as follows:
\begin{eqnarray}
&&\hspace*{-1.3cm}\frac{\partial}{\partial t}\rightarrow\frac{\partial}{\partial t}-\epsilon v_g\frac{\partial}{\partial \xi}+\epsilon^2\frac{\partial}{\partial\tau},
\label{3eq:13}\\
&&\hspace*{-1.3cm}\frac{\partial}{\partial x}\rightarrow\frac{\partial}{\partial x}+\epsilon\frac{\partial}{\partial \xi}.
\label{3eq:14}
\end{eqnarray}
Now, by substituting the Eqs. \eqref{3eq:8}-\eqref{3eq:14} into
Eqs. \eqref{3eq:1}, \eqref{3eq:2}, and Eq. \eqref{3eq:7}, and
collecting the terms containing $\epsilon$, the first order
($m=1$ with $l=1$) equations can be expressed as
\begin{eqnarray}
&&\hspace*{-1.3cm}ku_{i1}^{(1)}=\omega n_{i1}^{(1)},
\label{3eq:15}\\
&&\hspace*{-1.3cm}k\phi_1^{(1)}+k\alpha n_{i1}^{(1)}=\omega u_{i1}^{(1)},
\label{3eq:16}\\
&&\hspace*{-1.3cm}n_{i1}^{(1)}=k^2\phi_1^{(1)}+A_7\phi_1^{(1)},
\label{3eq:17}\
\end{eqnarray}
these equations reduce to
\begin{eqnarray}
&&\hspace*{-1.3cm}n_{i1}^{(1)}=\frac{k^2}{S}\phi_1^{(1)},
\label{3eq:18}\\
&&\hspace*{-1.3cm}u_{i1}^{(1)}=\frac{k \omega}{S}\phi_1^{(1)},
\label{3eq:19}
\end{eqnarray}
where $S= \omega^2-\alpha k^2$. We thus obtain the dispersion
relation for IAWs
\begin{eqnarray}
&&\hspace*{-1.3cm}\omega^2=\alpha k^2+\frac{k^2}{A_7+k^2}.
\label{3eq:20}
\end{eqnarray}
The second order ($m=2$ with $l=1$) equations are given by
\begin{eqnarray}
&&\hspace*{-1.3cm}n_{i1}^{(2)}=\frac{k^2}{S}\phi_1^{(2)} +\frac{2ik\omega(V_g k-\omega)}{S^2} \frac{\partial\phi_1^{(1)}}{\partial\xi},
\label{3eq:21}\\
&&\hspace*{-1.3cm}u_{i1}^{(2)}=\frac{k \omega}{S}\phi_1^{(2)}+\frac{i(V_g k-\omega)(\omega^2+ k^2\alpha)}{S^2} \frac{\partial\phi_1^{(1)}}{\partial\xi}, \label{3eq:22}
\end{eqnarray}
\begin{figure}[t!]
\centering
\includegraphics[width=70mm]{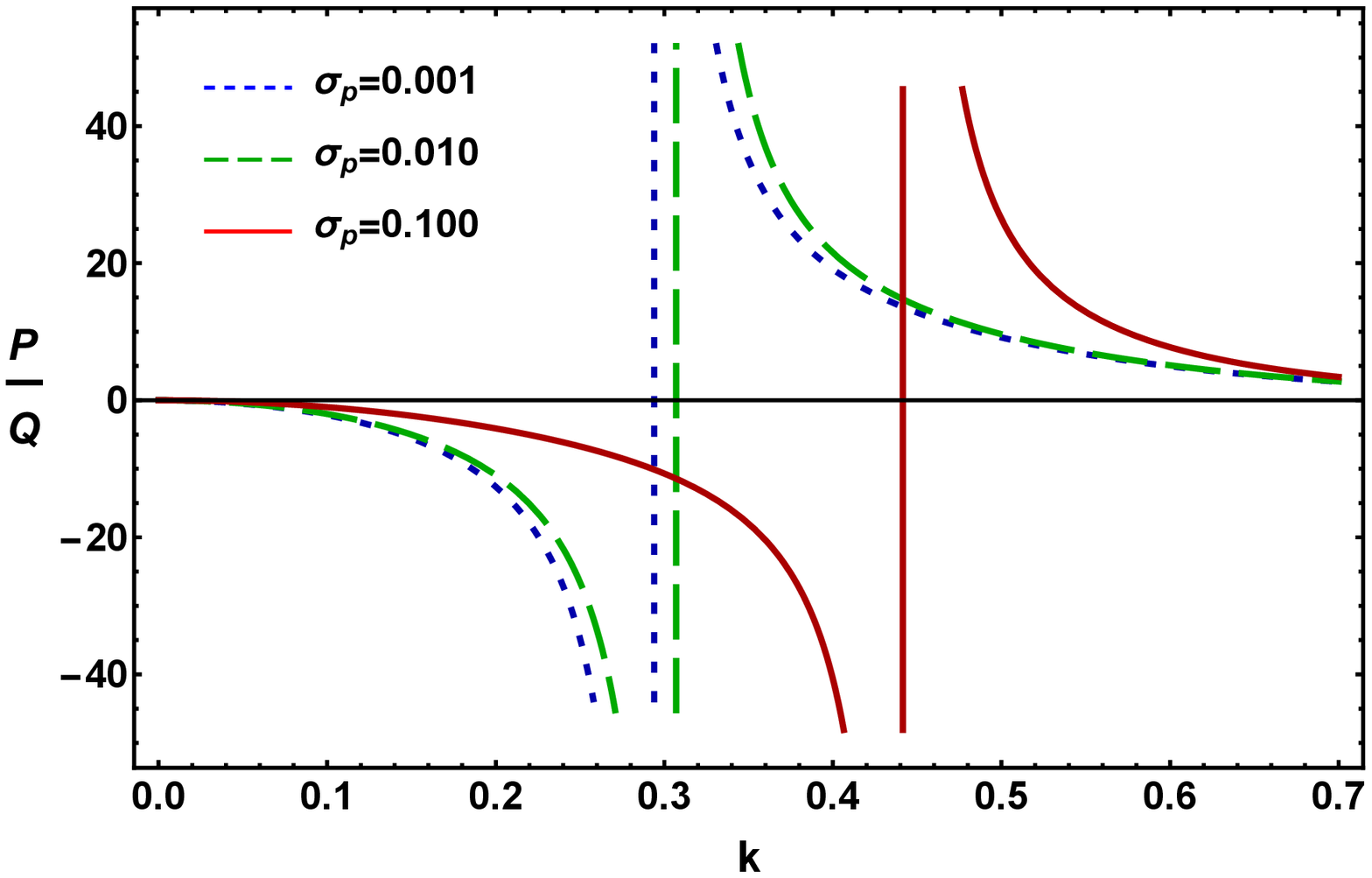}
\caption{The variation of $P/Q$ with $k$ for different values of
$\sigma_p$ when $\alpha=0.04$, $\gamma_c=0.5$, $\gamma_h=0.8$, $\sigma_h=0.01$,  $q_h=1.3$, and $q_p=1.5$.}
\label{3Fig:F2}
\end{figure}
with the compatibility condition
\begin{eqnarray}
&&\hspace*{-1.3cm}V_g=\frac{\partial \omega }{\partial k}=\frac{\omega^2-S^2}{\omega k}.
\label{3eq:23}
\end{eqnarray}
The coefficients of $\epsilon$ for $m=2$ and $l=2$ provide the second order harmonic amplitudes
which are found to be proportional to $|\phi_1^{(1)}|^2$
\begin{eqnarray}
&&\hspace*{-1.3cm}n_{i2}^{(2)}=A_{10}|\phi_1^{(1)}|^2,
\label{3eq:24}\\
&&\hspace*{-1.3cm}u_{i2}^{(2)}=A_{11} |\phi_1^{(1)}|^2,
\label{3eq:25}\\
&&\hspace*{-1.3cm}\phi_{2}^{(2)}=A_{12} |\phi_1^{(1)}|^2,
\label{3eq:26}\
\end{eqnarray}
where
\begin{eqnarray}
&&\hspace*{-1.3cm}A_{10}=\frac{\alpha k^6+ 3\omega^2 k^4+2A_{12}S^2k^2}{2S^3},~~A_{11}=\frac{\omega A_{10}S^2-\omega k^4}{kS^2},
~~A_{12}=\frac{\alpha k^6+3\omega^2 k^4-2A_8S^3}{2S^3(4k^2+A_7)-2k^2S^2}.
\nonumber\
\end{eqnarray}
Now, we consider the expression for ($m=3$ with $l=0$) and ($m=2$ with $l=0$),
which leads the zeroth harmonic modes. Thus, we obtain
\begin{eqnarray}
&&\hspace*{-1.3cm}n_{i0}^{(2)}=A_{13}|\phi_1^{(1)}|^2,
\label{3eq:27}\\
&&\hspace*{-1.3cm}u_{i0}^{(2)}=A_{14}|\phi_1^{(1)}|^2,
\label{2eq:28}\\
&&\hspace*{-1.3cm}\phi_0^{(2)}=A_{15} |\phi_1^{(1)}|^2,
\label{3eq:29}\
\end{eqnarray}
where
\begin{eqnarray}
&&\hspace*{-1.3cm}A_{13}=\frac{2\omega V_g k^3+\alpha k^4+\omega^2k^2+A_{15} S^2}{S^2(V_g^2-\alpha)},~~~~A_{14}=\frac{A_{13}V_gS^2-2\omega k^3}{S^2},
\nonumber\\
&&\hspace*{-0.3cm}A_{15}=\frac{2\omega V_g k^3+\alpha k^4+\omega^2k^2-2A_8S^2(V_g^2-\alpha)}{A_7S^2(V_g^2-\alpha)-S^2}.
\nonumber\
\end{eqnarray}
Finally, the third harmonic modes ($m=3$) and ($l=1$), with the help of
Eqs. \eqref{3eq:18}-\eqref{3eq:29}, give a set of equations, which can be
reduced to the following NLSE:
\begin{eqnarray}
&&\hspace*{-1.3cm}i\frac{\partial\Phi}{\partial\tau}+P\frac{\partial^2\Phi}{\partial\xi^2}+Q|\Phi|^2\Phi=0,
\label{3eq:30}
\end{eqnarray}
where $\Phi=\phi_1^{(1)}$ for simplicity. In Eq. \eqref{3eq:30}, $P$ is the dispersion coefficient which can be written as
\begin{eqnarray}
&&\hspace*{-1.3cm}P=\frac{V_g\alpha^2
k^5-3V_gk\omega^4+4\alpha k^2\omega^3-4\omega\alpha^2k^4+2\alpha V_g\omega^2k^3}{2\omega^2k^2},
\nonumber\
\end{eqnarray}
and $Q$ is the nonlinear coefficient which can be written as
\begin{eqnarray}
&&\hspace*{-1.3cm}Q=\frac{2A_8S^2(A_{12} + A_{15}) + 3A_9S^2
-(\omega^2k^2+\alpha k^4)(A_{10}+A_{13})-2\omega k^3(A_{11}+A_{14})}{2 \omega k^2},
\nonumber\
\end{eqnarray}
The space and time evolution of the IAWs in a FCPM are directly governed by the
coefficients $P$ and $Q$, and indirectly governed by different plasma parameters
such as $\alpha$, $\gamma_c$, $\gamma_h$, $\sigma_h$, $\sigma_p$, $q_h$, $q_p$,
and $k$, etc. Thus, these plasma parameters can significantly modify the stability conditions
of IAWs in a FCPM.
\begin{figure}[t!]
\centering
\includegraphics[width=70mm]{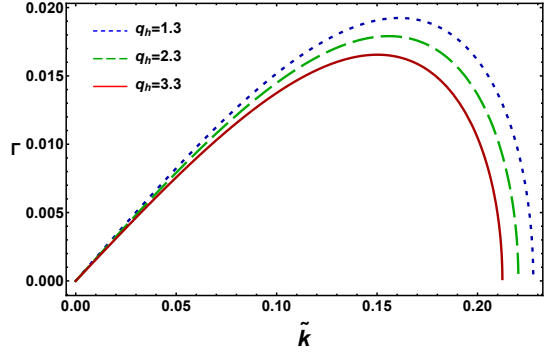}
\caption{The variation of $\Gamma$ against $\widetilde{k}$ for different values of $q_h$ when
$k=0.5$, $\phi_0=0.5$, $\alpha=0.04$, $\gamma_c=0.5$, $\gamma_h=0.8$, $\sigma_h=0.01$, $\sigma_p=0.01$,  and $q_p=1.5$.}
\label{3Fig:F3}
\end{figure}
\begin{figure}[t!]
\centering
\includegraphics[width=70mm]{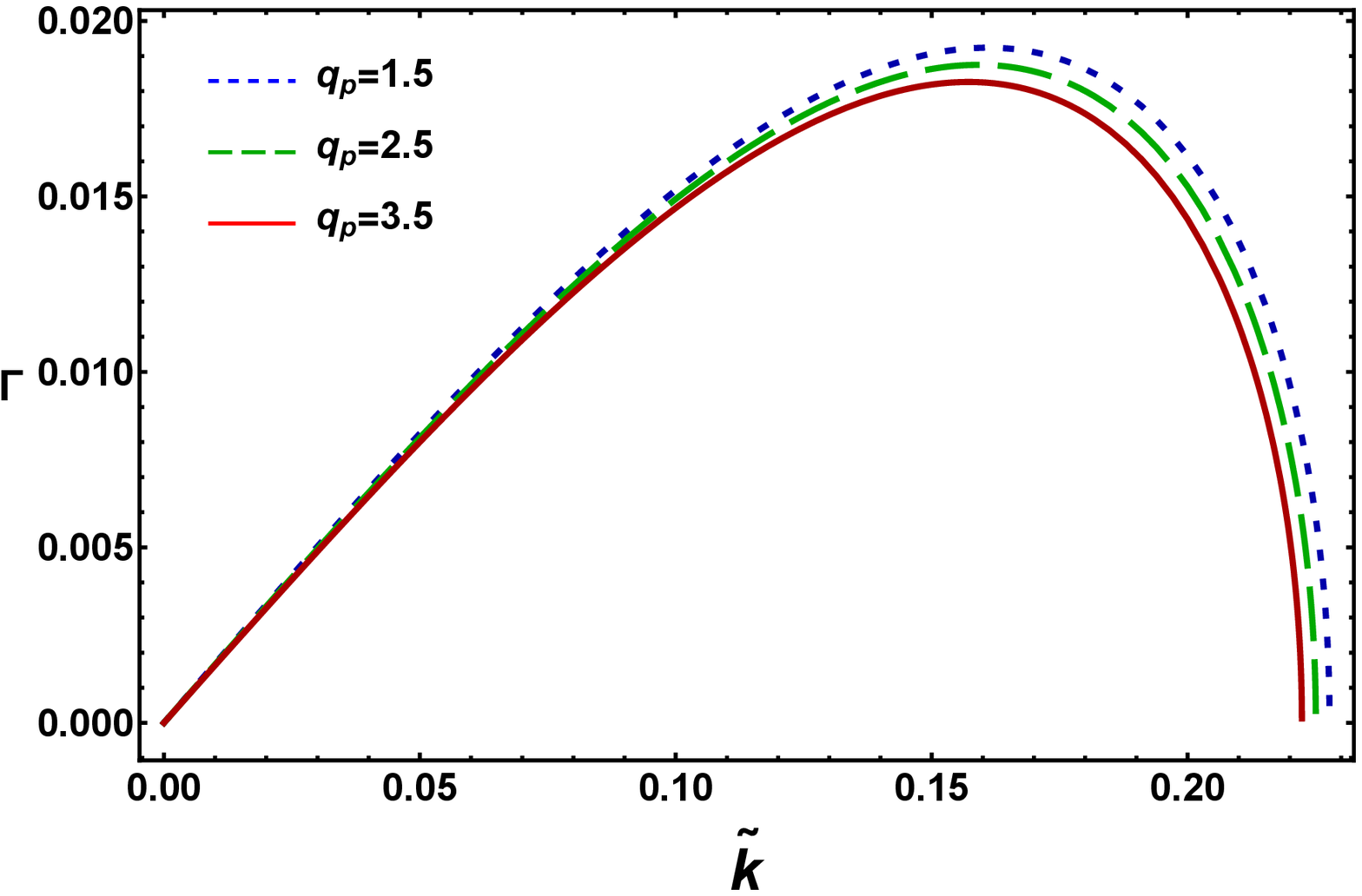}
\caption{The variation of $\Gamma$ against
$\widetilde{k}$ for different values of $q_p$ when
$k=0.5$, $\phi_0=0.5$, $\alpha$=$0.04$, $\gamma_c=0.5$, $\gamma_h=0.8$, $\sigma_h$=$0.01$, $\sigma_p$=$0.01$, and $q_h=1.3$.}
\label{3Fig:F4}
\end{figure}
\begin{figure}[t!]
\centering
\includegraphics[width=70mm]{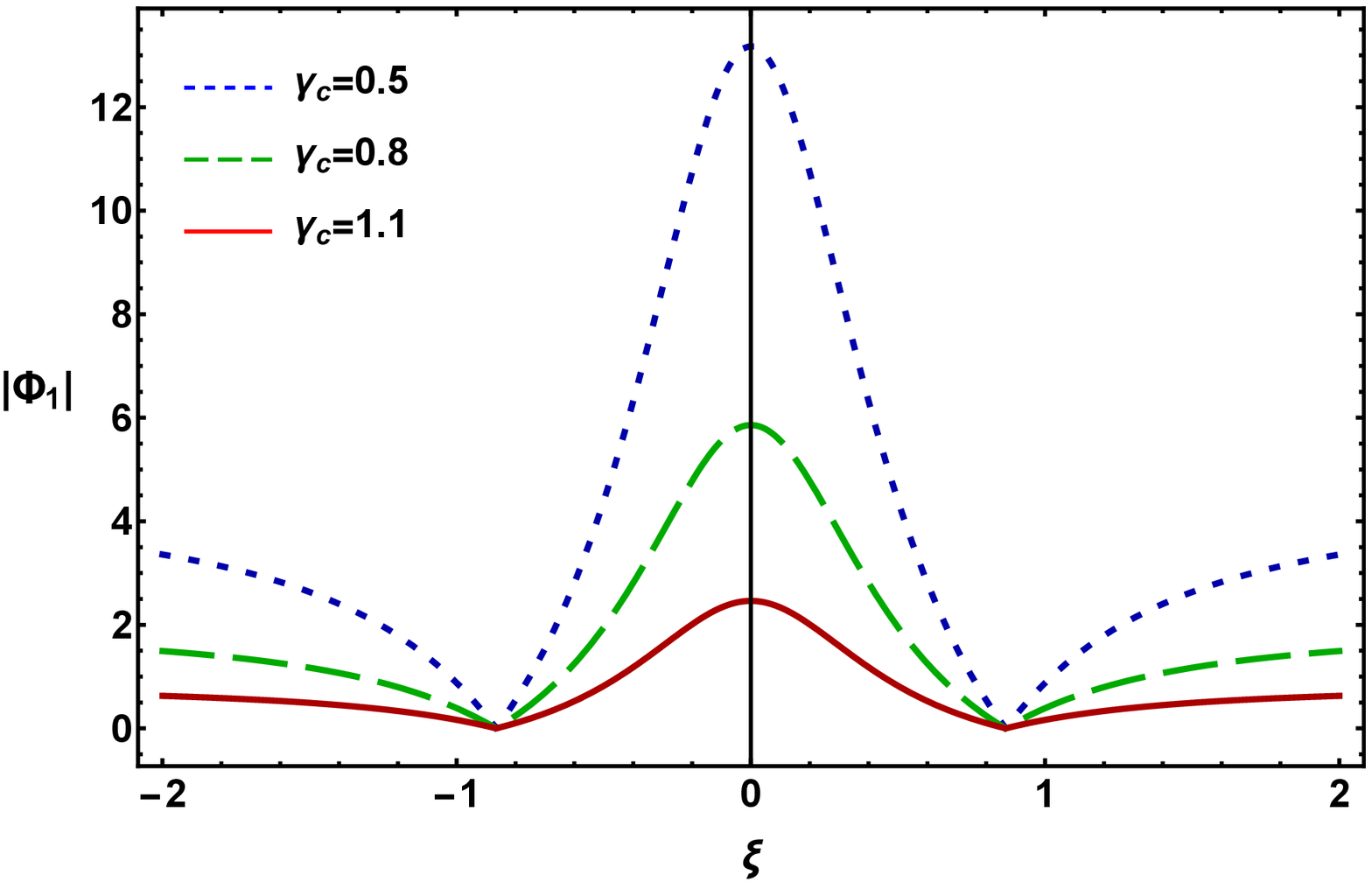}
\caption{The variation of the $|\phi|$ with $\xi$ for different
values of $\gamma_c$ when $k=0.5$, $\tau=0$, $\alpha=0.04$, $\gamma_h=0.8$, $\sigma_h=0.01$, $\sigma_p=0.01$, $q_h=1.3$, and $q_p=1.5$.}
\label{3Fig:F5}
\end{figure}
\begin{figure}[t!]
\centering
\includegraphics[width=70mm]{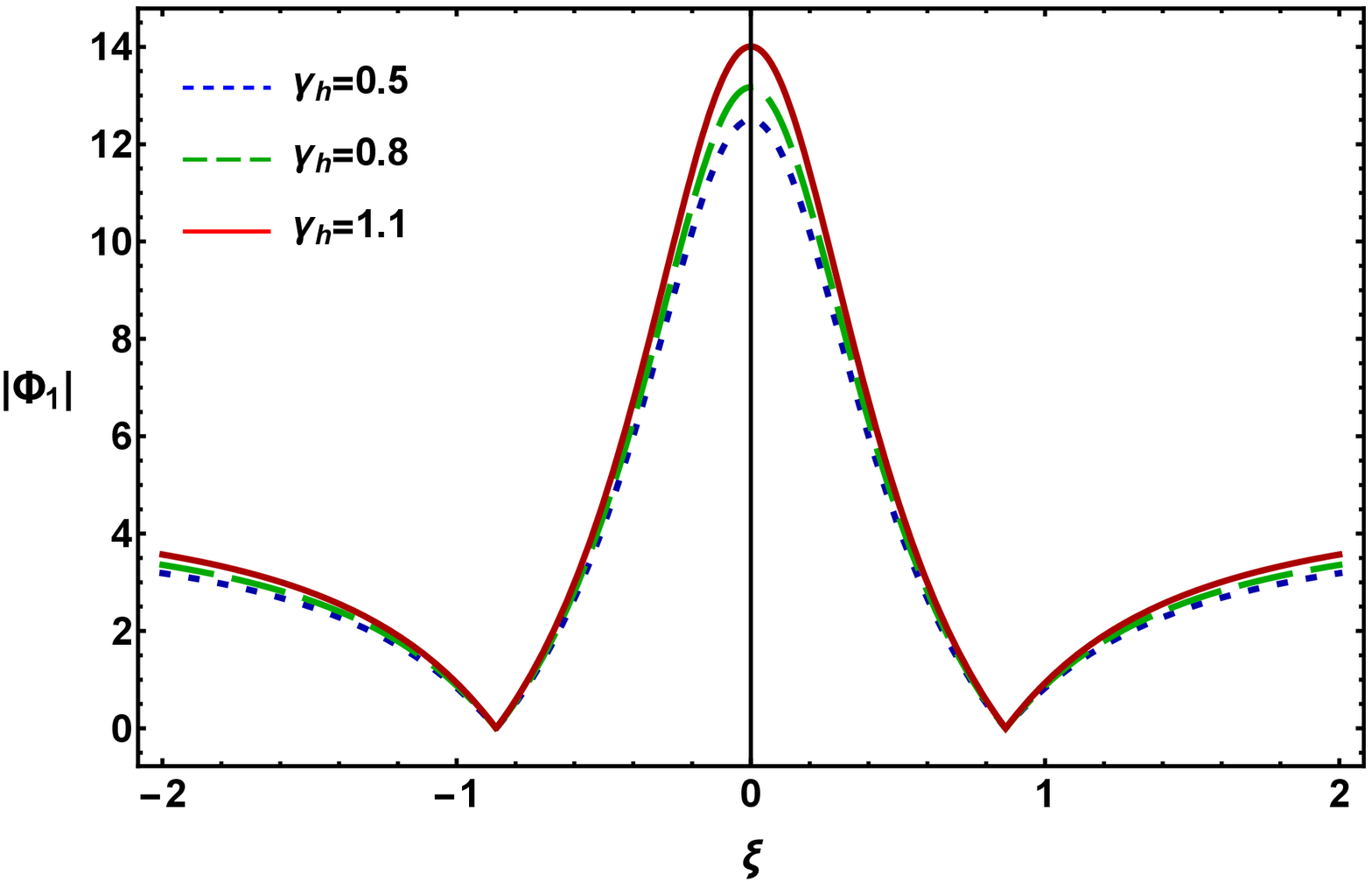}
\caption{The variation of the $|\phi|$ with $\xi$ for different
values of $\gamma_h$ when $k=0.5$, $\tau=0$, $\alpha=0.04$,  $\gamma_c=0.5$, $\sigma_h=0.01$, $\sigma_p=0.01$, $q_h$=$1.3$, and $q_p=1.5$.}
\label{3Fig:F6}
\end{figure}
\begin{figure}[t!]
\centering
\includegraphics[width=120mm]{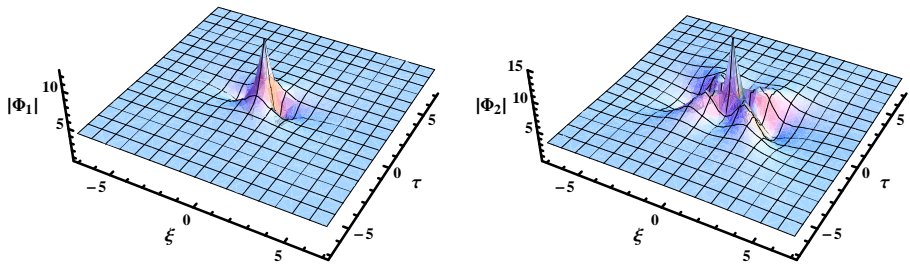}
\caption{Profile of the first-order rational solution (left
panel) and second-order rational solution (right panel) of NLSE at
$k=0.5$.}
\label{3Fig:F7}
\end{figure}
\begin{figure}[t!]
\centering
\includegraphics[width=70mm]{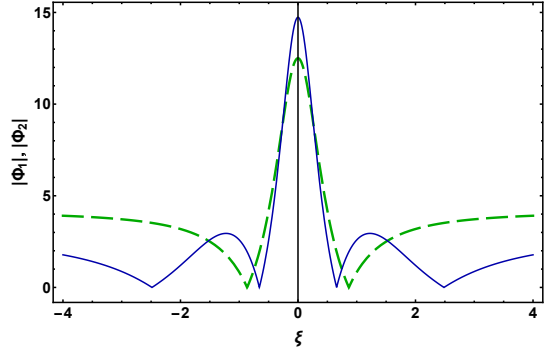}
\caption{The variation of first-order (dashed green curve) and
second-order (solid blue curve) rational solutions of NLSE at
$k=0.5$ and $\tau=0$.}
\label{3Fig:F8}
\end{figure}
\section{Modulational instability and Rogue waves}
\label{3sec:Modulational instability and Rogue waves}
To study the MI of IAWs, we consider the linear solution of the Eq.
\eqref{3eq:30} in the form $\Phi=\widetilde{\Phi}e^{iQ|\widetilde{\Phi}|^2\tau}$+c.c.,
where $\widetilde{\Phi}=\widetilde{\Phi}_0+\epsilon\widetilde{\Phi}_1$
and $\widetilde{\Phi}_1=\widetilde{\Phi}_{1,0}e^{i(\widetilde{k}\xi-\widetilde{\omega}{
\tau})}+c.c$. We note that the amplitude depends on the frequency, and that the perturbed
wave number $\widetilde{k}$ and frequency $\widetilde{\omega}$ which are different from $k$ and
$\omega$. Now, substituting these into Eq. \eqref{3eq:30}, one can easily obtain the following
nonlinear dispersion relation \cite{Chowdhury2019,Kourakis2005,Sultana2011}
\begin{eqnarray}
&&\hspace*{-1cm}\widetilde{\omega}^2=P^2\widetilde{k}^2\Big(\widetilde{k}^2-\frac{2|\widetilde{\Phi}_0|^2}{P/Q}\Big).
\label{3eq:31}
\end{eqnarray}
It is observed here that the ratio $P/Q$ is negative (i.e.,
$P/Q<0$), the IAWs will be modulationally stable. On the other
hand, if the ratio $P/Q$ is positive (i.e., $P/Q>0$), the IAWs
will be modulationally unstable \cite{Chowdhury2019,Kourakis2005,Sultana2011,Schamel2002,Fedele2002}.
It is obvious from Eq.
\eqref{3eq:31} that the IAWs becomes modulationally unstable
when $\widetilde{k}_c>\widetilde{k}$ in the regime $P/Q>0$, where
$\widetilde{k}_c = \sqrt{2(Q/P)}{|\widetilde{\Phi}_0|}$. The
growth rate $\Gamma$ of the modulationally unstable IAWs is
given by
\begin{eqnarray}
&&\hspace*{-1cm}
\Gamma=|P|\widetilde{k}^2\sqrt{\frac{\widetilde{k}_c^2}{\widetilde{k}^2}-1}.
\label{3eq:32}
\end{eqnarray}
The NLSE \eqref{3eq:30} has a variety of rational solutions, among them
there is a hierarchy of rational solution that are localized in
both the $\xi$ and $\tau$ variables.  The first-order rational solution of
Eq. \eqref{3eq:30} can be written as \cite{Ankiewiez2009,Guo2013,Guo2014,Yan2010}
\begin{eqnarray}
&&\hspace*{-1.3cm}\Phi_1 (\xi,
\tau)=\sqrt{\frac{2P}{Q}}\Big[\frac{4+16 i\tau P}{1+4 \xi^2 +
16\tau^2 P^2}-1\Big] \mbox{exp} (2i\tau P).
\label{3eq:33}
\end{eqnarray}
The interaction of the two or more first-order RWs can generate higher-order RWs
which has a more complicated nonlinear structure. The second-order rational
solution of Eq. \eqref{3eq:30} can be written as \cite{Ankiewiez2009,Guo2013,Guo2014,Yan2010}
\begin{eqnarray}
&&\hspace*{-1.3cm}\Phi_2 (\xi,\tau)=\sqrt{\frac{P}{Q}}\Big[1+\frac{G_2(\xi,\tau)+iM_2(\xi,\tau)}{D_2(\xi,\tau)}\Big]\mbox{exp} (i\tau P),
\label{3eq:34}
\end{eqnarray}
where
\begin{eqnarray}
&&\hspace*{-0.7cm}G_2(\xi,\tau)=\frac{3}{8}-6(P\xi\tau)^2-10(P\tau)^4-\frac{3\xi^2}{2}-9(P\tau)^2-\frac{\xi^4}{2},
\nonumber\\
&&\hspace*{-0.7cm}M_2(\xi,\tau)=-P\tau\Big[\xi^4+4(P\xi\tau)^2+4(P\tau)^4-3\xi^2+2(P\tau)^2-\frac{15}{4}\Big],
\nonumber\\
&&\hspace*{-0.7cm}D_2(\xi,\tau)=\frac{\xi^6}{12}+\frac{\xi^4(P\tau)^2}{2}+\xi^2(P\tau)^4+\frac{2(P\tau)^6}{3}+\frac{\xi^4}{8}
+\frac{9(P\tau)^4}{2}-\frac{3(P\xi\tau)^2}{2}+\frac{9\xi^2}{16}+\frac{33(P\tau)^2}{8}+\frac{3}{32}.
\nonumber\
\end{eqnarray}
The Eqs. \eqref{3eq:33} and \eqref{3eq:34} represent the
profile of the first and second order IARWs associated with the IAWs in the
modulationally unstable parametric regime (i.e., $P/Q>0$), respectively.
We have numerically analyzed the first and second order IARWs in Figs. \ref{3Fig:F5}-\ref{3Fig:F8}.
\section{Results and discussion}
\label{3sec:Results and discussion}
Now, we would like to numerically analyze the stability conditions
of the IAWs in presence of cold electrons following Maxwellian distribution, and hot
electrons and positrons featuring $q$-distribution. The existence of two temperature
electrons with distinct temperature and number density can be found in Saturn's
magnetosphere \cite{Panwar2014,Kourakis2003,Alinejad2014,Baluku2011,Baluku2012,Schippers2008},
Auroral plasma \cite{Temerin1982,Bostrom1988}, Earth's magnetosphere \cite{Gaffey1976,Ghosha1997},
tandem mirror experiments \cite{Kesner1985}, rf-heated plasma \cite{Nishida1986}, and
sputtering magnetron plasma \cite{Sheridan1991}, etc. The Saturn's magnetosphere has three
regions: the inner magnetosphere ($R\leq9R_s$), intermediate magnetosphere ($9R_s<R<13R_s$),
and outer  magnetosphere ($\geq13R_s$), where $R_S\approx60,300$ km is the radius of Saturn.
The components of the inner magnetosphere of Saturn are $N^+$, $O^+$, $OH^+$,  $H_2O^+$, and neutral objects \cite{Krupp2005}, etc.
Schippers \textit{et al.} \cite{Schippers2008} analysed the CAPS/ELS and MIMI/LEMMS data from
the Cassini spacecraft orbiting Saturn over a range of $5.4-20R_S$ which can be found from Table 1.
\begin{table}
\centering
    \caption{Parameter values derived from Schippers \textit{et al.} \cite{Baluku2011,Schippers2008} corresponding to Saturn's Magnetosphere}
     \begin{tabular}{lllll}
        \toprule
        $R$~($R_s$) & $T_c$ (eV) & $T_h$ (eV) & $n_c$ ($cm^{-3}$) & $n_h$ ($cm^{-3}$) \\
        \midrule
        5.40    & 1.8        & 300      & 10.5       & 0.02    \\
        6.30    & 2.0        & 400      & 10.5       & 0.01    \\
        9.80    & 8.0        & 1100     & 2.50        & 0.07    \\
        12.0    & 6.0        & 1200     & 1.00        & 0.11    \\
        13.1    & 10.2       & 1000     & 0.21       & 0.18    \\
        14.0    & 30         & 900      & 0.15       & 0.10    \\
        15.2    & 70         & 900      & 0.25       & 0.10    \\
        17.8    & 28         & 1000     & 0.15       & 0.07    \\
        \bottomrule
     \end{tabular}
\end{table}
A number of authors numerically analyzed the effects of two distinct temperature (hot and cold)
electrons following iso-thermal \cite{Rehman2016,Baboolal1989} or
non-thermal \cite{Shahmansouri2013,Shalini2015,Panwar2014,Kourakis2003,Alinejad2014,Baluku2012,Baluku2011}
distribution on the dynamics of space \cite{Rehman2016,Shahmansouri2013,Shalini2015,Panwar2014,Kourakis2003,Alinejad2014,Baluku2012,Baluku2011} and
laboratory \cite{Ghosha1997,Nishida1986,Sheridan1991,Baboolal1989} plasma system under these assumptions: $T_h>T_c$ and
$n_{h0}>n_{c0}$ \cite{Rehman2016,Shalini2015,Panwar2014,Kourakis2003,Alinejad2014,Baluku2012,Baluku2011,Ghosha1997,Nishida1986,Baboolal1989}
or $n_{h0}=n_{c0}$ \cite{Kourakis2003,Baluku2012,Baluku2011} or
$n_{h0}<n_{c0}$ \cite{Shahmansouri2013,Panwar2014,Alinejad2014,Baluku2012,Baluku2011,Sheridan1991,Baboolal1989}.
The parameters $q_h,~q_p$ are the non-extensive parameter describing the degree
of non-extensivity, i.e., $q_h,~q_p=1$ corresponds to Maxwellian distribution,
whereas $q_h,~q_p<1$ refers to the super-extensivity, and
the opposite condition $q_h,~q_p>1$ refers to the sub-extensivity.
This means that in the dynamics of electrons and positrons,
all the forces (including the force leading to annihilation of
electrons and positrons \cite{Kourakis2006,Khalejahi2006}) except the forces arising from
electrostatic wave potential, thermal pressure of electrons
and positrons, and deviation from Maxwellian to non-extensive
$q$-distribution have been neglected. Therefore, in our present investigation,
we have considered for our numerical analysis that $T_h=T_p=(10-1000)T_c$,
$T_i=0.1T_c$ \cite{Baluku2012,Nishida1986}, $Z=1-20$ \cite{Shahmansouri2013,Shalini2015,Panwar2014,Kourakis2003,Alinejad2014,Baluku2012,Baluku2011},
$n_{h0}>n_{c0}$, $n_{h0}=n_{c0}$, $n_{h0}<n_{c0}$, and small fraction of positrons.

The variation of $P/Q$ with $k$ for different values of $\sigma_h$
and $\sigma_p$ can be seen in Figs. \ref{3Fig:F1} and  \ref{3Fig:F2}, respectively, and these figures
can highlight the effects of temperature of the hot electron and positron as well as cold electron
species on the modulationally stable and unstable parametric regimes of IAWs in FCPM.
It is clear from these figures that (i) both stable and unstable parametric regimes
are allowed by the FCPM; (ii) the $k_c$ increases with the increase in the value of both $\sigma_h$ and $\sigma_p$;
(iii) the physics of this result is that the nonlinearity of the plasma medium increases with the increase of both hot electron
and positron temperature for constant temperature of the cold electrons, and this would lead
the IAWs become unstable for small values of $k$ as well as allows to generate
the first and second order IARWs in the modulationally unstable parametric regime (i.e., $k>k_c$).

We have numerically analyzed Eq. \eqref{3eq:32} in  Figs. \ref{3Fig:F3} and \ref{3Fig:F4}
to observed that how the nonlinearity as well as the growth rate of the IAWs changes with $\widetilde{k}$
for different values of the non-extensivity of hot electrons (via $q_h$)
and hot positrons (via $q_p$), and it is obvious from these figures that
an increase in the value of the $q_h$ (in Fig. \ref{3Fig:F3}) or $q_p$ (in Fig. \ref{3Fig:F4})
does not only cause to decrease the nonlinearity of the FCPM but also causes to decrease
the maximum value of the growth rate. The physics of the result is that the distribution
function with $q<1$, compared with the Maxwellian one ($q=1$)
indicates the system with more super-thermal particles (super-extensivity) whereas the $q$-distribution
with $q>1$ is suitable for plasma containing a large number low-speed particles (sub-extensivity).
This means that our FCPM has large number low-speed particles which reduce the nonlinearity as well as
the maximum value of the growth rate with $q_h$ and $q_p$.

Figure \ref{3Fig:F5} and \ref{3Fig:F6} indicate how the nonlinearity of FCPM as well as
the configuration of the IARWs associated with IAWs in the modulationally unstable parametric
regime (i.e., $P/Q>0$) changes with the charge state of positive ion and also the number
density of cold and hot electrons, and warm ions. The amplitude and the width of the IARWs
decrease with an increase in the value of the cold electron number density for a constant
value of the  charge state and number density of the warm ions (via $\gamma_c$ and can be
seen from Fig. \ref{3Fig:F5}). On the other hand, the existence of large amount of hot
electrons increases the amplitude and width of the IARWs associated with IAWS when other
plasma parameters remain constant (via $\gamma_h$ and can be seen from Fig. \ref{3Fig:F6}).
These two interesting phenomena may be explained in physical framework as follows: an
increase in cold (hot) electron number density could shrink (enhance) the nonlinearity of the FCPM and
disperse (concentrate) its energy which makes the amplitude and width of the IARWs shorter
and narrower (taller and wider).

The time evolution and the comparison of the first and second order IARW
associated with IAW in the modulationally unstable parametric regime can be seen from
Figs. \ref{3Fig:F7} and \ref{3Fig:F8}, respectively. Figure \ref{3Fig:F8} indicates the comparison of
the first and second order IARW solutions at $\tau=0$, and it is clear from this figure
that (a) the second-order IARW has double structures compared with the first-order IARW;
(b) the amplitude of the second-order IARW is always greater than the amplitude of the
first-order IARW; (c) the potential profile of the second-order IARW becomes more
spiky (i.e., the taller amplitude and narrower width) than the first-order IARW;
(d) the second (first) order IARW has four (two) zeros symmetrically located
on the $\xi$-axis; (e) the second (first) order IARW has three (one) local maxima.
\section{Conclusion}
\label{3sec:Conclusion}
We have considered a more general and realistic four component plasma model,
and have investigated the stable and unstable parametric regimes, which can be recognized
by the sign of the coefficients $P$ and $Q$ of NLSE, of IAWs.
The $k_c$ value, which divides the stable and unstable parametric regimes of IAWs,
totally depends on the temperature of hot and cold electrons. An increase in the
value of the $q_h$ or $q_p$ does not only cause to decrease the nonlinearity of
the FCPM but also causes to decrease the maximum value of the growth rate.
The numerical analysis has also shown that the amplitude and width of the IARWs
increase with an increase in the value of hot electron number density while decrease with
an increase in the value of cold electron number when other plasma parameters remain constant.
Finally, the finding of our present investigation may be applicable in explaining the formation
of the IARWs in Saturn's magnetosphere \cite{Panwar2014,Kourakis2003,Alinejad2014,Baluku2011,Baluku2012,Schippers2008},
Auroral plasma \cite{Temerin1982,Bostrom1988}, Earth's magnetosphere \cite{Gaffey1976,Ghosha1997},
tandem mirror experiments \cite{Kesner1985}, rf-heated plasma \cite{Nishida1986}, and
sputtering magnetron plasma \cite{Sheridan1991}, etc.


\begin{thebibliography}{99}

\bibitem{Rehman2016} M.A. Rehman, M.K. Mishra, Phys. Plasmas \textbf{23} (2016) 012302.

\bibitem{Shahmansouri2013} M. Shahmansouri, H. Alinejad, Phys. plasmas \textbf{24} (2017) 113701.

\bibitem{Shalini2015} Shalini, N.S. Saini, A.P. Misra, Phys. Plasmas \textbf{22} (2015) 092124.

\bibitem{Panwar2014} A. Panwar, C.M. Ryu, A.S. Bains, Phys. Plasmas  \textbf{21} (2014) 122105.

\bibitem{Kourakis2003} I. Kourakis, P.K. Shukla, J. Phys. A Math. Gen. \textbf{36} (2003) 11901.

\bibitem{Alinejad2014} H. Alinejad, M. Mahdavi, M. Shahmansouri, Astrophys. Space Sci. \textbf{352} (2014) 571.

\bibitem{Baluku2012} T.K. Baluku, M.A. Hellberg, Vacuum \textbf{147} (2018) 31.

\bibitem{Baluku2011} T.K. Baluku, M.A. Hellberg, R.L. Mace, J. Geophys. Res. \textbf{116} (2011) A04227.

\bibitem{Schippers2008}P. Schippers, M. Blanc, \textit{et al.}, J. Geophys. Res. \textbf{113} (2008) A07208.

\bibitem{Sittler1983} E.C. Sittler,  K.W. Ogilvie, J.D. Scudder, J. Geophys. Res. \textbf{88} (1983) 8847.

\bibitem{Barbosa1993} D.D. Barbosa, W.S. Kurth, J. Geophys. Res. \textbf{98} (1993) 9351.

\bibitem{Young2005}D.T. Young, J.J. Berthelier, \textit{et al.}, Science \textbf{307} (2005) 1262.

\bibitem{Ahmed2018}N. Ahmed, A. Mannan, N.A. Chowdhury, A.A. Mamun, Chaos \textbf{28} (2018) 123107.

\bibitem{Khondaker2019}S. Khondaker, N.A. Chowdhury, A. Mannan, A.A. Mamun, arXiv:1809.09312.

\bibitem{Chowdhury2017}N. A. Chowdhury, A. Mannan, M.M. Hasan, A.A. Mamun, Chaos \textbf{27} (2017) 093105.

\bibitem{Rahman2018a}M.H. Rahman, N.A. Chowdhury, A. Mannan, \textit{et al.}, Chinese J. Phys. \textbf{56} (2018) 2061.

\bibitem{Jahan2019}S. Jahan, N.A. Chowdhury, A. Mannan, A.A. Mamun, Commun. Theor. Phys. \textbf{71} (2019) 327.

\bibitem{Rahman2018b}M.H. Rahman, A. Mannan, N.A. Chowdhury, A.A. Mamun, Phys. Plasmas \textbf{25} (2018) 102118.

\bibitem{Chowdhury2019}N.A. Chowdhury, A. Mannan, M.M. Hasan, and A.A. Mamun, Plasma Phys. Rep. \textbf{45} (2019) 459.

\bibitem{Kourakis2005} I. Kourakis, P.K. Sukla, Nonlinear Proc. Geophys. \textbf{12} (2005) 407.

\bibitem{Sultana2011} S. Sultana, I. Kourakis, Plasma Phys. Control. Fusion \textbf{53} (2011) 045003.

\bibitem{Schamel2002} R. Fedele, H. Schamel, Eur. Phys. J. D \textbf{73} (2019) 177.

\bibitem{Fedele2002} R. Fedele, Phys. Scr. \textbf{65} (2002) 502.

\bibitem{Ankiewiez2009}A. Ankiewicz, P.A. Clarkson, N. Akhmediev, J. Phys. A \textbf{43} (2010) 12002.

\bibitem{Guo2013}S. Guo, L. Mei, W. Shi, Contrib. Plasma Phys. \textbf{58} (2018) 870.

\bibitem{Guo2014}S. Guo, L. Mei,  Phys. Plasmas \textbf{21} (2014) 112303.

\bibitem{Yan2010}Z. Yan,  Commun. Theor. Phys. \textbf{71} (2019) 1017.

\bibitem{Temerin1982}M. Temerin, K. Cerny, W. Lotko, F.S. Mozer, Phys. Rev. Lett. \textbf{48} (1982) 1175.

\bibitem{Bostrom1988}R. Bostrom, G. Gustafsson, \textit{et al.}, Phys. Rev. Lett. \textbf{61} (1988) 82.

\bibitem{Gaffey1976}J.D. Gaffey, R.E. LaQuey, J. Geophys. Res. \textbf{81} (1976) 595.

\bibitem{Ghosha1997}S.S. Ghosha, A.N.S. Iyengar, Phys. Plasmas \textbf{4} (1997) 3204.

\bibitem{Kesner1985}J. Kesner, Nucl. Fusion \textbf{25} (1985) 275.

\bibitem{Nishida1986} Y. Nishida, T. Nagasawa, Phys. Fluids \textbf{29} (1986) 345.

\bibitem{Sheridan1991} T. E. Sheridan, M.J. Goeckner, J. Goree, J. Vacuum Sci. Tech. A \textbf{9} (1991) 688.

\bibitem{Baboolal1989} S. Baboolal, R. Bharuthram, M.A. Hellberg, J. Plasma Phys. \textbf{41} (1989) 341.

\bibitem{Krupp2005} N. Krupp, A. Lagg, J. Woch, \textit{et al.}, Geophys. Res. Lett. \textbf{32} (2005) L20S03.

\bibitem{Kourakis2006} I. Kourakis, A. Esfandyari-Khalejahi, \textit{et al.}, Phys. Plasmas \textbf{13} (2006) 052117.

\bibitem{Khalejahi2006}A. Esfandyari-Khalejahi, I. Kourakis, \textit{et al.}, J. Phys. A \textbf{39} (2006) 13817.

\end{thebibliography}
\end{document}